\documentclass[twocolumn,twoside,slac]{revtex4}
\usepackage{graphicx}
\usepackage{fancyhdr}
\begin{document}

\title{Reconstruction of electrons with the Gaussian-sum filter in the
  CMS tracker at LHC}

%

\author{W.~Adam, R.~Fr{\"u}hwirth}
\affiliation{Institute for High-Energy Physics, Austrian Academy of
  Sciences, Vienna, Austria} 
\author{A.~Strandlie}
\affiliation{CERN, Geneva, Switzerland}
\author{T.~Todorov}
\affiliation{IReS, Strasbourg, France}

\begin{abstract}
    The bremsstrahlung energy loss distribution of electrons propagating
    in matter is highly non Gaussian. Because the Kalman filter relies
    solely on Gaussian probability density functions, it might not be an
    optimal reconstruction algorithm for electron tracks.
    A Gaussian-sum filter (GSF) algorithm for electron track reconstruction in the 
    CMS tracker has therefore been developed. The basic idea is to model the
    bremsstrahlung energy loss distribution by a Gaussian mixture rather than a
    single Gaussian. It is shown that the GSF is able to improve the momentum 
    resolution of electrons compared to the standard Kalman filter. 
    The momentum resolution and the quality of the estimated error are
    studied with various types of mixture models of the energy loss distribution.

\end{abstract}

\maketitle

\thispagestyle{fancy}


\section{Introduction}
Modern track detectors based on semiconductor technologies 
contain larger amounts of material than gaseous detector types,
partially due to the detector elements themselves and partially due to
additional material required for on-sensor electronics, power, cooling,
and mechanical support. A precise modelling of  material effects in
track reconstruction is therefore necessary to obtain
the best estimates of the track parameters. Such material
effects are particularly relevant for the 
reconstruction of electrons which, in addition to ionization energy
loss and multiple Coulomb scattering, suffer from large energy losses
due to bremsstrahlung.

\begin{figure}[hbtp]
  \begin{center}
    \resizebox{6.75cm}{!}{\includegraphics{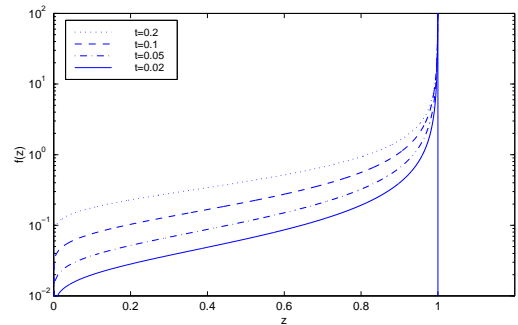}}
    \caption{Probability density function $f(z)$ for different
      thickness values.}\    
    \label{fig:BHpdf}
  \end{center}
\end{figure}

A well-known model of the bremsstrahlung energy loss is due to Bethe and 
Heitler~\cite{BetheHeitler34}. In this model, the probability density
function (PDF), $f(z)$, of the energy loss of an electron is
\begin{equation}
  f(z) = \frac{\left[- \ln z \right]^{c-1}}{\Gamma (c)},
\end{equation}
where $c = t / \ln 2$, $t$ is the thickness of material traversed by
the electron (in units of radiation length), and $z$ is the fraction
of energy remaining after the material layer is traversed.
The probability of a given fractional energy loss is assumed to be
independent of the energy of the incoming particle. 
This PDF is shown in Fig.~\ref{fig:BHpdf} for different thickness values.

The baseline for track reconstruction in the CMS tracker is the Kalman
filter~\cite{Rudi87}.
Throughout the filter tracks are described by a five-dimensional state
vector, containing the information about the
momentum, the direction and the position at some reference surface.
The material effects are currently assumed to be concentrated in the
active elements of the detector layers.
In this context
the optimal treatment of radiative energy loss  
is to correct the momentum with the mean value of energy loss
and to increase the variance of the momentum by adding the variance of
the energy loss distribution.
This procedure should ensure unbiased estimates of the track parameters
and of the associated uncertainties~\cite{Stampfer94}. The Kalman
filter is a linear least-squares estimator, and is proved to be optimal
only when all probability
densities encountered during the track reconstruction procedure are Gaussian. The
implicit assumption of approximating the Bethe-Heitler distribution with a single
Gaussian is quite crude. It is therefore plausible that a non-linear estimator
which takes the actual shape of the distribution into account can do better.

A non-linear generalization of the Kalman filter (KF), the {\em Gaussian-sum filter 
(GSF)}~\cite{Rudi97, Rudi98}, has therefore been implemented in the reconstruction 
software of the CMS tracker~\cite{ORCA}. 
In the GSF the distributions of all state vectors are Gaussian
mixtures, i.e. weighted sums of Gaussians instead of single Gaussians. 
The algorithm is therefore appropriate if the
probability densities involved in track reconstruction can be
adequately described by Gaussian 
mixtures. The basic idea of the present work is to approximate the
Bethe-Heitler distribution as a Gaussian mixture rather than a single
Gaussian, in which the different components of the mixture model
different degrees of hardness of the bremsstrahlung in the layer under
consideration. The resulting estimator resembles a set of Kalman filters
running in parallel, where each Kalman filter corresponds to one of
the components of the mixture describing the distribution of the state vector. 

\section{Approximating the fractional energy loss distribution}
\begin{figure}[hbtp]
  \begin{center}
    \resizebox{6.75cm}{!}{\includegraphics{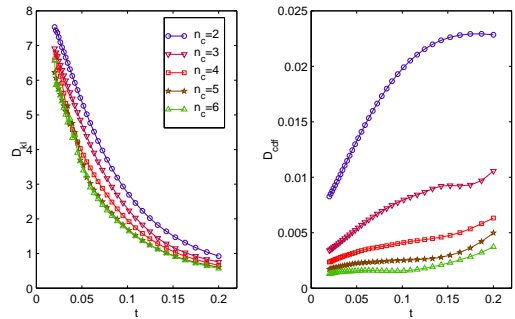}}
    \caption{The distances $D_{\mbox{{\scriptsize KL}}}$ and
      $D_{\mbox{{\scriptsize CDF}}}$ as a function of the thickness
      $t$, for CDF-mixtures, with different numbers of components.}\
    \label{fig:KLdist}
  \end{center}
\end{figure}

\begin{figure}[hbtp]
  \begin{center}
    \resizebox{6.75cm}{!}{\includegraphics{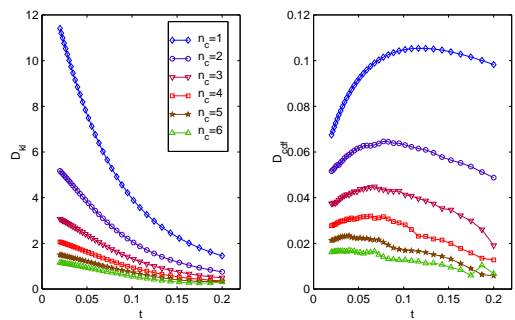}}
    \caption{The distances $D_{\mbox{{\scriptsize KL}}}$ and
      $D_{\mbox{{\scriptsize CDF}}}$ as a function of the thickness
      $t$, for KL-mixtures, with different numbers of components.}\
    \label{fig:CDFdist}
  \end{center}
\end{figure}

An important issue with the GSF reconstruction of electrons is to
obtain a good 
Gaussian-mixture approximation of the Bethe-Heitler distribution. The
parameters to be obtained are the weights, the mean values and the
variances of each of the components in the approximating mixture. The
parameters are determined by minimizing the following two
distances:
\begin{eqnarray}
  D_{\mbox{{\scriptsize CDF}}} & = & \int_{- \infty}^{\infty} \left| F(z) - G(z)
  \right| dz, \\
  \label{equation:KLdist}
  D_{\mbox{{\scriptsize KL}}} & = & \int_{- \infty}^{\infty} \ln \left[ f(z)/g(z)
  \right] f(z) dz, 
\end{eqnarray}
where $f(z)$ and $F(z)$ are the PDF and cumulative distribution
function (CDF) of the model distribution and $g(z)$ and $G(z)$ are the
PDF and CDF of the Gaussian mixture, respectively. The distance
$D_{\mbox{{\scriptsize KL}}}$ is the so-called Kullback-Leibler
distance between the model distribution and the mixture. 
Hereafter, the mixtures 
obtained by minimizing $D_{\mbox{{\scriptsize CDF}}}$ are called
CDF-mixtures, whereas the mixtures obtained by minimizing
$D_{\mbox{{\scriptsize KL}}}$ are called KL-mixtures. The
minimizations have been done independently on a set of discrete values
of $t$, ranging from 0.02 to 0.20.
Figures~\ref{fig:KLdist}
and~\ref{fig:CDFdist} show the resulting distances as a function of
thickness for a
varying number of components in the approximating mixture.
In order to obtain mixtures for arbitrary values of the thickness,
fifth-degree polynomials have been fitted to the parameters as a
function of $t$. 
Due to the fast access
to the parameters from the polynomials, the calculation of the mixture
is done on the fly during
reconstruction, using the effective thickness of a detector layer
from the knowledge of the incident angle of inclination.

\section{Reducing the number of components}
The approximation of energy loss by a Gaussian mixture amounts to a
convolution of this mixture with the current state, which in general
is also composed of several Gaussian components. 
The strict application of the GSF algorithm therefore quickly leads to a
prohibitively large number of components due to the combinatorics
involved each time a layer of material is traversed.

In a realistic implementation of the GSF the number of components must 
repeatedly be reduced to a predefined maximum. 
As little information as possible should be lost in this procedure.
Two strategies have been tested:
\begin{enumerate}
\item Only the {\it N} components with the largest weights are kept;
\item Components are merged into clusters, according to a given metric.
\end{enumerate}

The first option has the advantage of being computationally light, but 
it turns out to be inferior. Even the first two moments of the
estimated parameters are not described correctly.

In the second approach, the component with the largest weight is
merged with the one closest to it, and this procedure is repeated
until the required number of components is reached. The results below
have been obtained by using the Kullback-Leibler distance -- defined in
Equation~(\ref{equation:KLdist}) -- as a measure of distance. 

\section{Results from simulated tracks in the CMS tracker}
\begin{figure}[hbtp]
  \begin{center}
    \resizebox{6.75cm}{!}{\includegraphics{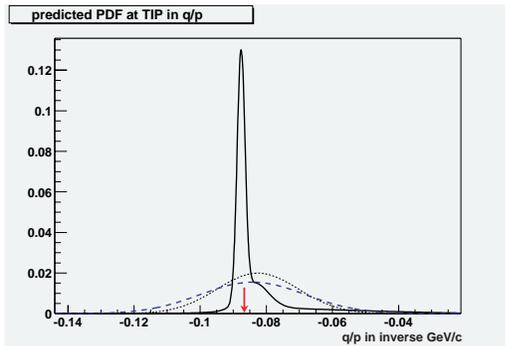}}
    \caption{Estimated $q/p$ of one single track for the GSF (solid),
      the KF (dashed) and
      the combined GSF state (dotted). The combined GSF state refers
      to the first and the second moments of the GSF estimate, here
      visualized as a single Gaussian. The arrow denotes the true
      value of $q/p$. It can be seen that the estimated PDF of the GSF
      is a non-Gaussian function.}\    
    \label{fig:qpSingleTrack}
  \end{center}
\end{figure}

First, results from the reconstruction of data originating from a simplified
simulation are shown. In this simulation multiple scattering and
ionization energy loss are turned off, all the material is
concentrated on the detector units, and the exact amount of material used in the
simulation is known by the reconstruction program. Single electron
tracks with $p_T = 10 $ GeV/$c$ have been simulated for absolute
values of $\eta$ less than 1.0 . Reconstructed
hits have been collected using the knowledge of the associated
simulated hits, so no pattern recognition 
has been involved. The following results all refer to the
quantity $q/p$ (charge over absolute value of the momentum) recorded at
the point of closest approach to the vertex in 
the transverse plane -- the transverse impact point (TIP) -- after a
fit going from the outside towards the inside of the tracker. 
Figure~\ref{fig:qpSingleTrack} 
shows an example
of the estimated $q/p$ for one single track, both for the KF and for
the GSF.

\begin{figure}[hbtp]
  \begin{center}
    \includegraphics[width=6.75cm]{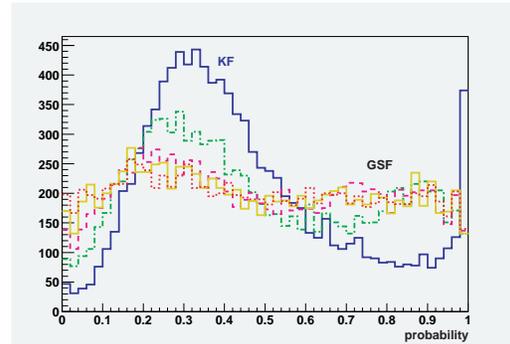}
    \caption{Probability distribution for the estimated $q/p$ for the KF
      (solid) and the GSF with a maximum of six (dashed-dotted),
      twelve (dashed), $18$ 
      (solid) and $36$ (dotted) components kept during reconstruction. In
      this case the same six-component CDF-mixture has been used both
      in the simulation of the 
      disturbance of the momentum in a detector unit and in
      reconstruction. Keeping 36 components yields estimates
      quite close to the correct distribution of the parameter.}\    
    \label{fig:pullProbsMixture}
  \end{center}
\end{figure}

\begin{figure}[hbtp]
  \begin{center}
    \includegraphics[width=6.75cm]{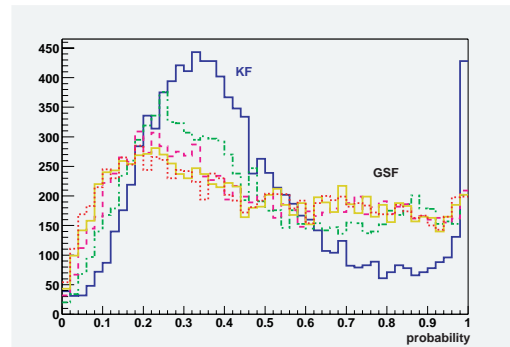}
    \caption{Probability distribution for the estimated $q/p$ for the KF
      (solid) and the GSF with a maximum of six (dashed-dotted),
      twelve (dashed), $18$ 
      (solid) and $36$ (dotted) components kept during reconstruction. The
      same six-component mixture as the one described in the caption
      of Fig.~\ref{fig:pullProbsMixture} has been used in
      reconstruction, but the simulation of the disturbance of the
      momentum in a detector unit has been done by sampling from the
      Bethe-Heitler distribution. The distributions for the GSF are
      seen to be less flat than those shown in
      Fig.~\ref{fig:pullProbsMixture}.}\   
    \label{fig:pullProbs}
  \end{center}
\end{figure}

Figures~\ref{fig:pullProbsMixture} and~\ref{fig:pullProbs} 
show probability distributions for the estimated $q/p$ of the KF and the
GSF with a varying maximum number of components kept during the reconstruction. Given the
estimated PDF (a single Gaussian for the KF, a Gaussian mixture for
the GSF), each entry in the histogram amounts to the integral from $-\infty$ to the
true value of $q/p$. If the estimated PDF is a correct description of
the real distribution of the parameter, the corresponding
histogram should be flat. 

\begin{figure}[hbtp]
  \begin{center}
    \resizebox{6.75cm}{!}{\includegraphics{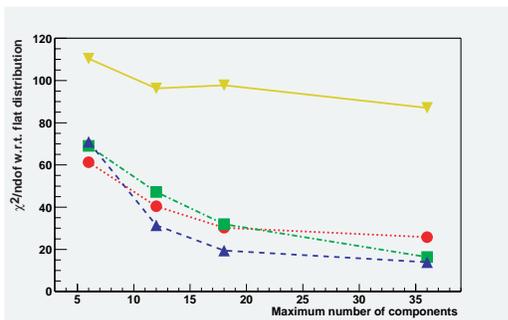}}
    \caption{Values of $\chi^2$ per bin of the probability distribution for CDF-mixtures with four
      (circles), five (squares) and six (triangles pointing upwards)
      components, as well as a KL-mixture 
      with six (triangles pointing downwards) components. The
      corresponding value for the KF is $146$.}\    
    \label{fig:probChi2}
  \end{center}
\end{figure}

The deviation from flatness can be quantified by the
$\chi^2$ of the difference between the probability distributions of
$q/p$ and the flat distribution. This $\chi^2$ per bin is shown in
Fig.~\ref{fig:probChi2}
for a set of different mixtures as a function of the maximum number of
components kept. The CDF-mixtures are superior to the
KL-mixtures concerning the quality of the estimated $q/p$.
The main trend seems to be related to the maximum number
of components kept rather than the number of components in the mixture
describing the energy loss, even though the mixtures with five and six
components are best in the limit of keeping a large number of
components.

\begin{figure}[hbtp]
  \begin{center}
    \resizebox{6.75cm}{!}{\includegraphics{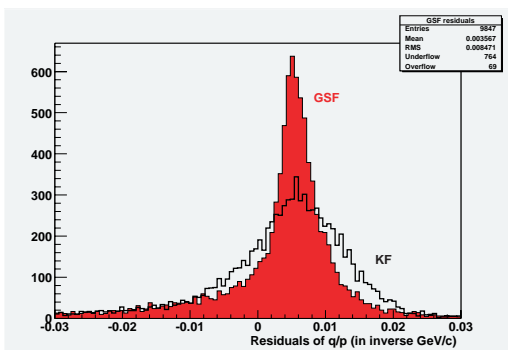}}
    \caption{Residuals of estimated $q/p$ with respect to the true
      value of the parameter for the GSF and the KF at the
      transverse impact point. A maximum number of twelve components
      has been kept during reconstruction. Long
      tails extending outside the limits of the histogram exist both
      for the KF and for the GSF. These tails are due to hard
      radiation in the innermost layers of the tracker.}\    
    \label{fig:qpPredTipBH}
  \end{center}
\end{figure}

\begin{figure}[hbtp]
  \begin{center}
    \resizebox{6.75cm}{!}{\includegraphics{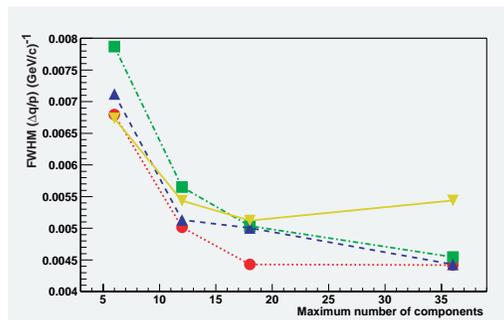}}
    \caption{Full-width at half-maximum for the GSF as a function of the
      maximum number of components kept during reconstruction. The
      corresponding value of the KF is $0.013$. }\    
    \label{fig:resFwhm}
  \end{center}
\end{figure}

\begin{figure}[hbtp]
  \begin{center}
    \resizebox{6.75cm}{!}{\includegraphics{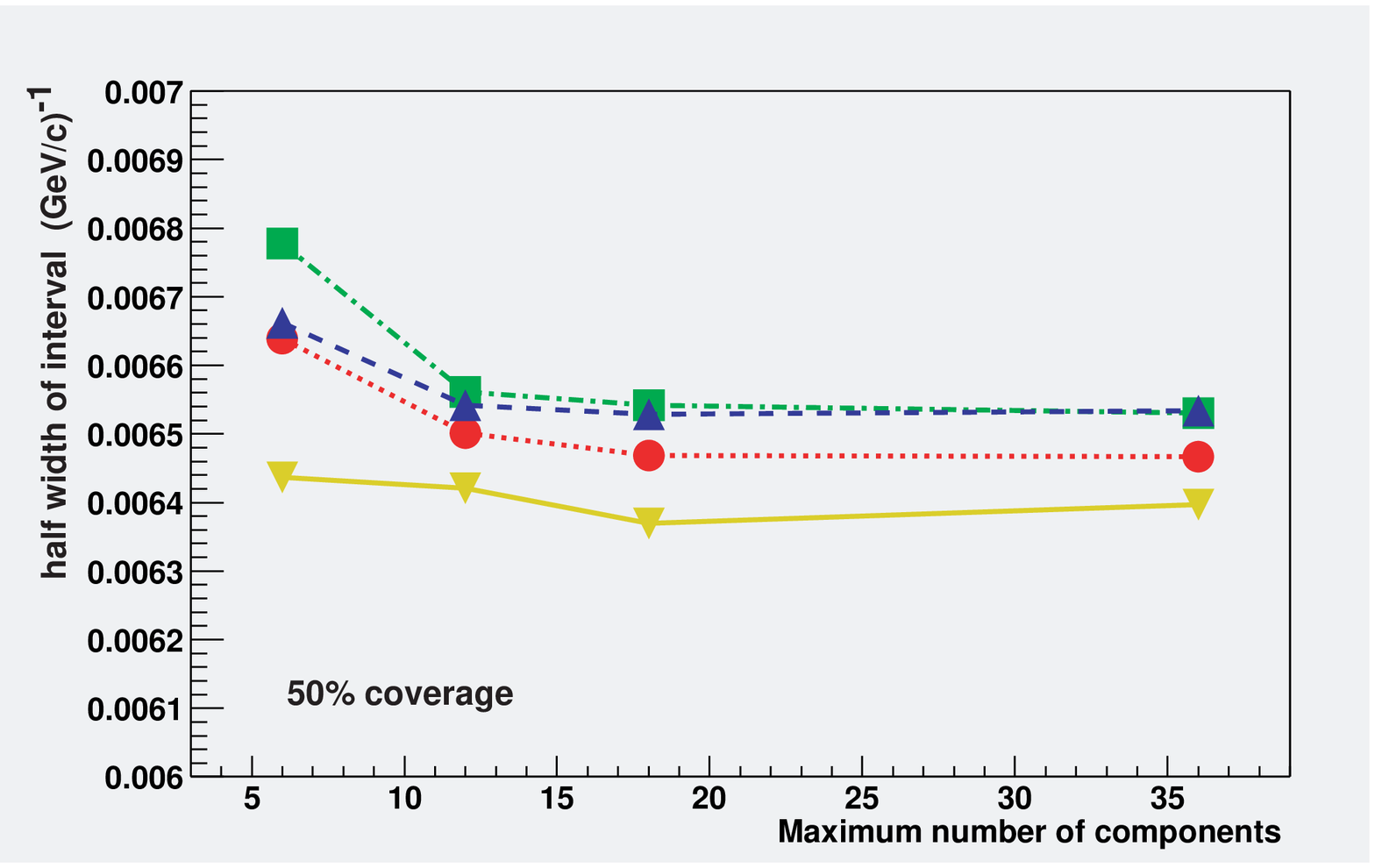}}
    \caption{Half-widths of the symmetric intervals covering 50\% of the
      distribution for the GSF as a function of the
      maximum number of components kept during reconstruction. The
      corresponding value of the KF is $0.0080$.}\    
    \label{fig:resQ5}
  \end{center}
\end{figure}

\begin{figure}[hbtp]
  \begin{center}
    \resizebox{6.75cm}{!}{\includegraphics{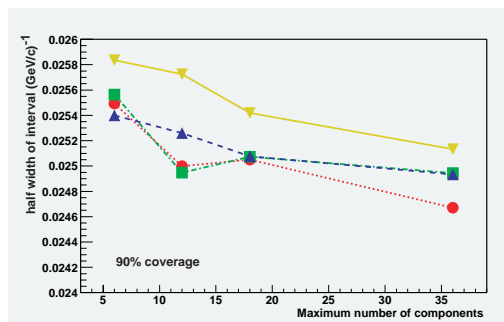}}
    \caption{Half-widths of the symmetric intervals covering 90\% of the
      distribution for the GSF as a function of the
      maximum number of components kept during reconstruction. The
      corresponding value of the KF is $0.0295$.}\    
    \label{fig:resQ9}
  \end{center}
\end{figure}

Figure~\ref{fig:qpPredTipBH}
shows the residuals of the estimated $q/p$ of the GSF and the KF
with respect to the true value of the parameter. The estimated $q/p$
for the GSF is the mean value of the state vector mixture, and the
mixture used for
this specific plot is a CDF-mixture with six components. In order to
quantify the difference between the GSF and the KF residuals, the
full-width at half-maximum (FWHM) and the half-width of intervals
covering 50\% and 90\% of the distribution have been considered. 
The covering intervals have been chosen to be symmetric about zero. 
The FWHM and the half-widths of the covering intervals are shown in
Figs.~\ref{fig:resFwhm},~\ref{fig:resQ5} and~\ref{fig:resQ9}.
The different flavours of the GSF in these figures are the same as
those described in the caption of Fig.~\ref{fig:probChi2}.


\begin{figure}[hbtp]
  \begin{center}
    \resizebox{6.75cm}{!}{\includegraphics{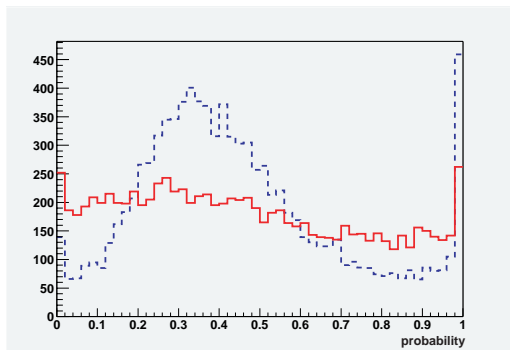}}
    \caption{Probability distribution for the estimated $q/p$ for the KF
      (dashed) and the GSF (solid). The specific mixture used in the GSF
      is a CDF-mixture with six components. A maximum number of twelve
      components has been kept during reconstruction. The
      reconstruction algorithms have been run on tracks from a full
      simulation.}\    
    \label{fig:pullProbsFull}
  \end{center}
\end{figure}

The GSF and the KF have also been run on tracks from a full simulation
using the official CMS simulation program~\cite{CMSIM}. The $p_T$ and
the $\eta$ range are the same as in the simplified simulation, but the
amount and spatial distribution of the material are different. Probability
distributions of the estimated $q/p$ for the GSF and the 
KF are shown in Fig.~\ref{fig:pullProbsFull}.
The probability distribution of the GSF exhibits no large deviation from flatness,
indicating that the estimated PDF of $q/p$ describes reasonably well
the actual PDF of $q/p$. This observation
is all the more remarkable since, with the full simulation, the energy loss is
not generated by the simple Bethe-Heitler model, and neither the exact
amount nor the exact location of the material are known to the GSF.

\begin{figure}[hbtp]
  \begin{center}
    \resizebox{6.75cm}{!}{\includegraphics{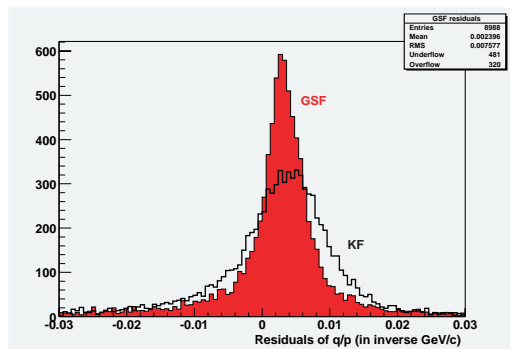}}
    \caption{Residuals of the estimated $q/p$ with respect to the true
      value at the transverse impact point for the KF and the GSF. The
      reconstruction algorithms have been run on tracks from a full
      simulation.}\    
    \label{fig:qpPredTipFull}
  \end{center}
\end{figure}

\begin{figure}[hbtp]
  \begin{center}
    \resizebox{6.75cm}{!}{\includegraphics{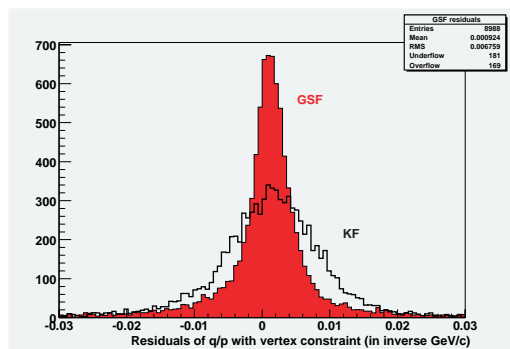}}
    \caption{Residuals of the estimated $q/p$ with respect to the true
      value at the transverse impact point for the KF and the GSF. A
      vertex constraint has been included in the fits. The
      reconstruction algorithms have been run on tracks from a full
      simulation.}\    
    \label{fig:qpUpdTipFull}
  \end{center}
\end{figure}

The corresponding residuals of the estimated $q/p$ with respect to the
true value are shown in Figs.~\ref{fig:qpPredTipFull}
and~\ref{fig:qpUpdTipFull}.
The residuals shown in Fig.~\ref{fig:qpUpdTipFull} have been obtained
by including a vertex constraint in the fit. Such a constraint allows
the momentum to be measured in the innermost part of the track and thus gives
a handle on possible radiation in the first two layers.
The result of including this constraint is
a less skew distribution with the mode being moved closer
towards zero, and the amount of tracks in the tails is also
reduced. Even though the results from the full simulation
qualitatively seem to confirm those from the simplified simulation,
more studies are needed to understand the differences in detail.

\section{Conclusion}
The Gaussian-sum filter has been implemented in the CMS reconstruction
program. It has been validated with electron tracks with a simplified
simulation in which the energy loss distribution (Bethe-Heitler
model), the exact amount of 
material and its exact location are known to the reconstruction program.
It has been shown that the quality of the momentum estimate depends mainly
on the number of mixture components kept during reconstruction, and to
some extent also on the number of components in the mixture approximation
to the energy loss distribution. A comparison with the best linear
unbiased estimator, the Kalman filter, shows a clear improvement of the
momentum resolution. Remarkably, a similar improvement can be seen with
electron tracks from the full simulation, although in this case neither
the exact energy loss distribution nor the precise amount and location of
material are known to the reconstruction program. More systematic studies
with electrons from the full simulation are clearly needed, but it seems
safe to conclude that in electron reconstruction the Gaussian-sum filter
yields a substantial gain in precision as compared to the Kalman filter.

\end{document}